\documentclass[9pt,conference]{IEEEtran}
\usepackage{dcase2025}

\usepackage{soul}
\usepackage{url}
\usepackage{hyperref}
\usepackage[utf8]{inputenc}
\usepackage{graphicx}
\usepackage{amsmath, amssymb}
\usepackage{bbm}
\usepackage{amsthm}
\usepackage{booktabs}
\usepackage{algorithm}
\usepackage{algorithmic}
\usepackage{multicol}
\usepackage{multirow}
\usepackage[switch]{lineno}
\usepackage{acronym}
\usepackage{makecell}
\usepackage{xspace}
\usepackage{placeins}
\usepackage{sidecap}

\acrodef{iou}[IoU]{intersection over union}
\acrodef{sed}[SED]{sound event detection}
\acrodef{zf}[ZF]{zebra finch}
\acrodef{nms}[NMS]{non-maximal suppression}
\acrodefplural{zf}[ZFs]{zebra finches}

\newcommand{\method}{\texttt{Voxaboxen}\xspace}

\usepackage{bm} 

\usepackage{dcase2025,amsmath,graphicx,url,times,booktabs, tabularx}

\title{Robust detection of overlapping bioacoustic sound events}

\name{%
  \begin{tabular}{c}
  Louis Mahon\textsuperscript{1,2,*},
  Benjamin Hoffman\textsuperscript{1,*},
  Logan James\textsuperscript{1,3},
  Maddie Cusimano\textsuperscript{1},
  Masato Hagiwara\textsuperscript{1},\\
  \textit{Sarah C Woolley\textsuperscript{3},
  Felix Effenberger\textsuperscript{1},
  Sara Keen\textsuperscript{1},
  Jen-Yu Liu\textsuperscript{1},
  Olivier Pietquin\textsuperscript{1}}
  \end{tabular}
}

\address{%
\textsuperscript{1} Earth Species Project, USA \;
\textsuperscript{2} University of Edinburgh, UK \;
\textsuperscript{3} McGill University, CA\\
\textsuperscript{*} Equal contribution
}

\begin{document}

\maketitle

\begin{abstract}
We propose a method for accurately detecting bioacoustic sound events that is robust to overlapping events, a common issue in domains such as ethology, ecology and conservation. While standard methods employ a frame-based, multi-label approach, we introduce an onset-based detection method which we name \method. 
For each time window, \method predicts whether it contains the start of a vocalization and how long the vocalization is. It also does the same in reverse, predicting whether each window contains the end of a vocalization, and how long ago it started, and fuses the two sets of bounding boxes with a graph-matching algorithm. We also release a new dataset of temporally-strong labels of zebra finch vocalizations designed to have high overlap. Experiments on eight datasets, including our new dataset, show \method outperforms natural baselines and existing methods, and is robust to vocalization overlap.
\end{abstract}


\section{Introduction}
Detecting animal sounds is the foundation of bioacoustics research. In practice, these sounds often overlap, but identifying each individual acoustic unit is necessary for a diversity of tasks, including species recognition and population estimation, 
which can be critical for ecology and conversation \cite{stowell2022}.

When multiple individuals from a single species co-occur, the sounds they produce can overlap with each other, often with important functional consequences, 
e.g. in bats~\cite{Gillam2016}, zebra finches~\cite{Elie2011-sm}, frogs~\cite{Clulow2017-yu}, and elephants~\cite{Soltis2005-zi}. 
To understand these communication systems, large-scale identification of individual vocalizations, including accurate classification of overlapping sounds, is of critical importance. 

Motivated by this, we desire a \ac{sed} method that can predict the onset time, offset time, and class label (e.g., species label) for overlapping sound events. Commonly, \ac{sed} methods adopt a frame-based approach: for each time frame, for each class, predicting whether a sound of that class occurs in that frame~\cite{CakirCRNN,frameATST,rookognise,sebbs}, and merging consecutive frames with the same class into a single event.
This does not accommodate overlaps from the same class. 


To address this limitation, we propose a method we name \method, 
For each frame, \method makes a binary prediction as to whether it contains an event onset, plus a regression prediction for how long that event will last, and a class prediction (e.g. species label). This design choice 
means the duration of one predicted event can extend past the onset of a second event, thus allowing the model to predict overlapping vocalizations without them being merged.

To investigate how well \method deals with overlapping vocalizations, we introduce a new dataset of recordings of eight female \acp{zf} spontaneously interacting in a laboratory environment, annotated with onset and offset of each vocalization, and featuring a high degree of overlap. 
We also introduce a series of synthetic datasets, consisting of \ac{zf} vocalizations with a controlled overlap-to-vocalization ratio. 
We find that \method consistently outperforms alternatives, even in the presence of a high degree of overlap, on our new dataset as well as seven previously-published bioacoustics datasets.

Taken together, our results demonstrate the general effectiveness of \method for bioacoustic \ac{sed}, including for situations with overlapping vocalizations.
To democratize putting boxes around vocalizations, we open source the code for our model and new dataset.

To summarize, the contributions of this paper are as follows:
\begin{enumerate}
    \item introducing \method, and \ac{sed} model leveraging pretrained audio encoders, which can predict overlapping vocalizations; 
    \item releasing a new dataset, Overlapping Zebra Finch (OZF), specifically focused on overlapping vocalizations; 
    \item  experimental evaluation on a diverse set of eight datasets, showing SotA performance for \method.
\end{enumerate}

\section{Related Work}


In bioacoustics applications, \ac{sed} has typically been framed as a multi-label classification problem~\cite{stowell2022}, with temporal resolution ranging from tens of milliseconds~\cite{tweetynet, rookognise}, to multiple seconds~\cite{ghani2023global,robinson2025naturelm}. Recent post-processing techniques decouple event durations and detections~\cite{sebbs,tomoya2025}; but still use frame-based predictions and cannot handle within-class overlaps. 
Other approaches include matrix factorization algorithms~\cite{dessein2013real} 
or probabilistic models~\cite{stowell2015acoustic}.

Visual object detection methods such as Faster-RCNN~\cite{ren2016faster} can accommodate overlapping objects, and have occasionally been applied to bioacoustic \ac{sed}~\cite{zsebHok2019automatic}. CornerNet \cite{law2018cornernet} is an object detection method that, similar to \method, matches predicted boundaries into a single event, but differs in that it matches boxes based on feature similarity, which can be inaccurate for animal vocalizations, where highly stereotyped 
 events mean that different events can share very similar features. Our approach accounts for this by matching based on \ac{iou} instead.

Given an audio recording with a mixture of sound sources, source separation is the task of predicting the audio of the pre-mixture sounds. 
Prior work in bioacoustics~\cite{denton2022improving} has demonstrated the effectiveness of source separation for improving accuracy in downstream classification tasks. In our context, a source separation model could theoretically separate vocalizations from multiple individuals into different audio tracks, thus reducing the complexity of the audio passed to a downstream detection model. We investigate this type of approach as an alternative to \method. 

A related task is speaker diarization, which segments recordings of multiple speakers and assigns each segment to a speaker. 
Approaches typically assume a maximum number of speakers (e.g., two or four), and make use of the assumption that speakers can be re-identified by their vocal characteristics across multiple segments~\cite{bredin2021end}.
In contrast, we assume no maximum number of speakers, and do not expect to re-identify individuals within a recording. 

\section{Method} \label{sec:method}

\subsection{Bounding Box Regression}

Our method, which is architecture-agnostic, uses a frame-based audio encoder $\phi\colon \mathbb{R}^T \to \mathbb{R}^{T'\times F}$ to produce a sequence of latent vectors. Here $T$ is the original number of samples, $T'$ is the final number of frames, and $F$ is the feature dimension.
A final linear layer $h\colon \mathbb{R}^{F}\to \mathbb{R}^{2+C}$
makes three types of predictions, for each time frame: a prediction of the probability that an event starts in that frame, a prediction of the duration of the event (should it start in that frame), and a prediction of a class label (logits across $C$ classes).

Using gradient descent, we minimize the loss function $L=L_{det}+\lambda L_{reg}+\rho L_{cls}$, $\lambda,\rho\ge0$, which includes a detection term $L_{det}$, a regression term $L_{reg}$, and a classification term $L_{cls}$. The detection term is inspired by the penalty-reduced focal loss in~\cite{law2018cornernet}:
\begin{equation}
L_{det} = -\frac{1}{T} \sum_{t=1}^T \left\{
                \begin{array}{lr}
                  (1-\hat{p}_t)^\alpha \log \hat{p}_t &  p_t = 1\\
                  (1-p_t)^\beta \hat{p}_t^\alpha \log(1-\hat{p}_t) & p_t<1.
                \end{array}
              \right.
\label{loss_eqn}
\end{equation}
Here, $T$ is the duration in frames of the audio clip, and $\alpha, \beta$ are hyperparameters.
In~\eqref{loss_eqn}, the model's predicted detection probability at time $t$ is $\hat{p}_t$, and the target $p_t$ is obtained by smoothing each event onset with a Gaussian kernel and taking the maximum value at each frame, across all events (following~\cite{law2018cornernet}):
\begin{equation}
\label{smooth_eqn}
p_t = \max_{x \in \operatorname{Events}} \exp\left(-\frac{(t-\operatorname{Onset}(x))^2}{\operatorname{Dur}(x)^2/s}\right)\,.
\end{equation}
In~\eqref{smooth_eqn}, $\operatorname{Events}$ is the set of events in an audio clip, and for $x\in \operatorname{Events}$, $\operatorname{Onset}(x)$ and $\operatorname{Dur}(x)$ denote the onset time and duration of $x$, and $s$ is a hyperparameter.

The regression term $L_{reg}$ is L1 loss, applied only to frames in $\{\operatorname{Onset}(x) \mid x \in \operatorname{Events}\}$, i.e. frames where an event begins.
Similarly, the classification term $L_{cls}$ is a categorical cross-entropy loss, again applied only when an event begins.
At inference time, we apply a peak-finding algorithm to the time-series of detection probabilities. Detection peaks above a threshold become boxes, with duration and class prediction determined by the value of the regression and classification predictions at the peak. The detection threshold is swept (for computing metrics), or fixed as a hyperparameter; see Section~\ref{sec:experimental-evaluation}. Finally, we apply soft non-maximal suppression~\cite{bodla2017soft} to remove duplicate boxes. 

\subsection{Bidirectional Predictions}

One drawback of using these predicted boxes directly is the difficulty for the model in making accurate regression predictions. In preliminary experiments, we observed that both onset and duration predictions can be slightly inaccurate, meaning that the model sometimes correctly detects an event but the edges of the bounding box are slightly off where they should be. To reduce error in bounding box edges, we make a second set of \textit{backward} predictions which are the mirror image of the first (\textit{forward}) set. The backward predictions are a binary prediction for each frame as to whether it contains an offset, plus a regression for how long the event lasted. We then compute an optimal way to fuse the forward and backward predictions into a single set of predictions, by casting the problem as a maximal bipartite graph matching problem. The bipartite graph has all boxes as vertices. Forward and backward boxes are linked by an edge if their \ac{iou} exceeds a threshold. The Hopcroft-Karp-Karzonov algorithm~\cite{hopcroft1973n} computes the maximal matching sub-graph, and edge-linked box pairs are fused. The onset of the fused box is defined to be the midpoint of the onset of the forward box, with the offset minus duration of the backward box (and similarly for the offset of the fused box).

\section{Overlapping Zebra Finch Dataset (OZF)} \label{sec:OZF-dataset}


\subsection{OZF Real-World Portion}
We recorded 65 minutes (divided into 60-second files) of 8 adult ($>1$ year) female \acp{zf} 
 housed in a large group cage in a sound attenuating chamber (TRA Acoustics, Ontario, Canada). We continuously recorded using Audacity (3.3.3) through two omnidirectional microphones (Countryman, Menlo Park, CA) positioned above and to the side of the cage. Food and water were provided \textit{ad libitum} and all procedures were approved by the McGill University 
Animal Care and Use Committee in accordance with Canadian Council on Animal Care guidelines. 

Female \acp{zf} make short, discrete vocalizations of about 100ms, consisting of a flat or downward sweeping harmonic stack, with most energy located between 0.5 and 8 kHz. 
The recordings were divided among three annotators, who marked the onset and offset time of each vocalization using Raven Pro (Cornell Lab of Ornithology, v.1.6.5). Annotators covered 25 minutes each. One 5 minute section was annotated by all three, where the mean pairwise inter-annotator F1@0.5\ac{iou} of 93.5, and 78.1 on the subset that overlaps. 

Out of a total of 8504 vocalizations in the dataset, 1449 (17.04\%) overlap with at least one other. The total number of overlaps is slightly higher at 1463, as some can overlap more than one other. The number of vocalizations per 60 s file ranges from 19 to 245, with between 0 and 73 overlapping. The duration of silence per 60 s file ranges from 35.5 to 58.1 seconds. We observe a roughly linear relationship between the two.
The duration of each vocalization ranges from 3 ms to 350 ms and is strongly peaked around the mean of 109 ms. 

It is possible to show that, assuming independent vocalizations from each bird that can be modelled with a Poisson distribution, the expected number of pairwise overlaps is $d(n-1) (1 - 1/B - 1/n)$,
where $d$ is the ratio of total call durations to time window size, $n$ is the number of vocalizations and $B$ is the number of birds (details provided at project github). In our case, $B=8$, $n \approx 120$ and $d \approx 0.1$. Plugging in the values for $n$ and $d$ from each 60s file 
the average ratio of overlap to number of vocalizations should be 20.46\%, significantly above the 17.04\% we observe.
This is consistent with prior work showing evidence for turn-taking in female \acp{zf} ~\cite{benichov2016forebrain}.


\subsection{OZF Synthetic Portion}

We generate six additional, synthetic \ac{zf} datasets where in each 60-second file, the ratio of overlaps to number of calls is controlled. We add cropped denoised female \ac{zf} vocalizations, drawn randomly from a call database, to a background track (recorded using the same setup as the OZF dataset, but without vocalizations). 


There are 65 recordings in each synthetic dataset, each mimicking one 60-second source recording in OZF (same number of vocalizations and placed in the same train/val/test split). To construct one synthetic recording, we count the number of calls present in the OZF source recording and randomly draw an equal number from the call database, and add them sequentially to the background track, to achieve a target overlap-to-call ratio in $R \in \{0.0, 0.2, 0.4, 0.6, 0.8, 1.0\}$
When $R>0.2$, the first 25\% of calls are placed uniformly at random to reduce clumping of calls around one point. Otherwise, each call is placed to overlap an existing call if and only if the current overlap ratio is below $R$. This method achieves an overlap ratio within $0.005$ of $R$ for each choice of $R$. Call amplitude is chosen so that signal-to-noise ratio falls uniformly at random between -15dB and 0dB.

\begin{table*}[h!]
    \begin{center}
    \resizebox{0.9\textwidth}{!}{
\begin{tabular}{lrrrrrlll}
\toprule
Dataset & Cls & Dur (h) & Events & Avg dur (s) & Overlap \% & Type & Location & Taxa \\
\midrule
AnSet\cite{canas2023dataset} & 10 & 26.82 & 7807 & 6.23 & 0.61 & TPAM & Brazil & Anura \\
\midrule
BV10\cite{nolasco2023learning} & 1 & 10.00 & 9024 & 0.15 & 21.68 & TPAM & NY, USA & Passeriformes \\
\midrule
HawB\cite{amanda_navine_2022_7078499} & 9 & 50.88 & 55713 & 1.11 & 2.54 & TPAM & HI, USA & Aves \\
\midrule
HbW\cite{humpbackdataset} & 1 & 13.38 & 4776 & 0.99 & 5.07 & UPAM & N Pacific & M. novaeangliae \\
\midrule
Katy\cite{madhusudhana2024extensive} & 1 & 4.49 & 11961 & 0.17 & 12.57 & TPAM & Panama & Tettigoniidae \\
\midrule
MT\cite{nolasco2023learning} & 1 & 1.26 & 1294 & 0.15 & 0.08 & On-body & S Africa & S. suricatta \\
\midrule
Pow\cite{Chronister2021} & 6 & 6.42 & 9919 & 1.11 & 6.99 & TPAM & PA, USA & Passeriformes \\
\midrule
OZF & 1 & 1.08 & 8504 & 0.11 & 17.20 & Lab & Lab & T. castanotis \\
\bottomrule
\end{tabular}
}
    \end{center}
    \caption{Summary of datasets used to evaluate model performance. Terrestrial and underwater passive acoustic monitoring units are abbreviated as TPAM and UPAM, respectively. Avg dur (s): average duration of vocalizations. Overlap \%: Number of within-class overlaps divided by number of vocalizations}\label{tab:dataset_metadata}
\end{table*}

\section{Experimental Evaluation} \label{sec:experimental-evaluation}

\textbf{Implementation Details}
We first extract features from the raw audio using a backbone encoder, and then make the predictions described in Section \ref{sec:method} from the extracted features. The encoder converts input audio (mono, 16 kHz) to a frame-based representation, which is a sequence of latent vectors produced at 50 Hz (window size 10s, hop size 5s). For the main experiments, we use BEATs~\cite{chen2023beats} as a backbone encoder. BEATs is an encoder-only transformer, with 12 layers, hidden size 768 and 8 attention heads, pretrained on Audioset \cite{gemmeke2017audio}. In Section \ref{subsec:ablation-studies}, we explore different choices of backbone. The detection, regression and classification predictions are then each made using a linear layer. The loss function hyperparameters were fixed at $\alpha = 2$, $\beta = 4$, and $s=6$ following~\cite{law2018cornernet}. During training and inference, audio is divided into 10-second windows, with 5-second step size between windows.  Training lasts for 50 epochs, with the encoder frozen for the first 3 epochs. We use Adam with ams-grad, $\beta_1=0.9$, $\beta_2=0.999$, and a cosine annealing scheduler. For all models, we select a learning rate from \{1e-4, 3e-5, 1e-5\}, based on mean average precision @0.5\ac{iou} on the val set. 
We apply soft non-maximal suppression~\cite{bodla2017soft} with $\sigma=0.5$.

\textbf{Datasets}
In addition to our newly released OZF dataset, we evaluated \method using seven existing datasets (Table~\ref{tab:dataset_metadata}), 
selected for their taxonomic diversity: amphibians (AnuraSet), insects (Katydid), birds (BirdVox-10h, Hawaiian Birds, Powdermill), and mammals (Humpback, Meerkat).
The preprocessing steps we performed on these datasets is described at the project github. 
For Katy, BV10 and OZF, the events of interest were brief and, for Katy and BV10, often above the 8kHz Nyquist frequency assumed by several of the models we evaluated. For all models, we use a half-time version of BV10 and OZF, and a sixth-time version of Katy. This effectively increases the output frame rate to 100 Hz for BV10 and OZF, and 300 Hz for Katy. 
Initial experiments indicated that using these slowed-down versions dramatically improved performance.

\textbf{Evaluation} As a metric, we first match predicted events to true events as in \cite{nolasco2023learning}, only counting matches that exceed a certain \ac{iou} threshold. Then, we compute mean average precision (mAP)
using 1001 equally-sized intervals.
We report results for an \ac{iou} threshold of 0.5 and of 0.8.

\textbf{Comparison Models}
We compare the performance of \method to several frame-based methods. Three of these consist of a linear layer on top of a encoder-only transformer, initialized with pre-trained weights. The encoders are
Frame-ATST~\cite{frameATST} (25 Hz output frame rate, pretrained on AudioSet), BEATs~\cite{chen2023beats} (50 Hz, pretrained on AudioSet)  and BirdAVES~\cite{HagiwaraAVES}\footnote{\url{https://github.com/earthspecies/aves}} (50 Hz, pre-trained on animal sound datasets).
Outputs are median filtered, with kernel size (ks) 1, 3, 7, or 11, selected based on mean average precision @0.5IoU on the val set.  \looseness=-1 

As an additional frame-based method, we compare to a convolutional-recurrent neural network (CRNN)~\cite{CakirCRNN,tweetynet,rookognise}. Model inputs are log-mel spectrograms (256 mel bands), and the model consists of a 2d conv layer (ks=7, hidden size 64), mean-pooling in the frequency dimension (ks=2), two 2d residual blocks 
(ks=3), mean pooling in both directions (ks=2), and finally a bi-LSTM, with hidden size 1024. The weights are randomly initialized.

Finally, we compare to two existing object detection models coming from computer vision, Faster-RCNN~\cite{ren2016faster} (X-101 model checkpoint pretrained on MS COCO)\footnote{\url{https://github.com/facebookresearch/detectron2}} and SEDT \cite{ye2021sound}, an encoder-decoder transformer, adapted to detect 1d events from a spectrogram\footnote{\url{https://github.com/Anaesthesiaye/sound_event_detection_transformer}}.


We implement an additional baseline that uses pre-trained source-separation model, BirdMixIT model~\cite{denton2022improving}, to separate audio into four stems. These stems are fed into BirdAVES, followed by a linear classification layer fine-tuned on the original (non-separated) data. We combine the four sets of detections and apply soft non-maximal suppression to de-duplicate boxes. We evaluated this method on OZF-synthetic, but based on its performance there did not evaluate it on the other datasets.


\subsection{Main Results} \label{subsec:main-results}

\begin{table*}[h!]
\tiny
\centering
\resizebox{\textwidth}{!}{
\begin{tabular}{llcccccccc}
\toprule
Metric & Method & AnSet & BV10 & HawB & HbW & Katy & MT & Pow & OZF \\
\midrule
\multirow{7}{*}{mAP@0.5IoU} 
     & CRNN & 9.89 & 35.59 & 22.72 & 21.03 & 17.24 & 82.97 & 35.45 & 71.80 \\
     & Faster-RCNN &8.06&55.49&7.39&21.66&25.93& 84.22 &14.08& 90.20 \\
     & SEDT & 0.18 & 3.79 & 2.79 & 3.95 & 2.30 & 18.58 & 2.71 & 2.26 \\
     & Frame-ATST &14.87 & 40.62 & 32.19 & 33.62 & 17.88 & 87.58 & 45.42 & 73.48 \\
     & BEATs &15.71 & 48.01 & 35.37 & 37.13 & 20.12 & 86.08 & 50.32 & 77.94 \\
     & BirdAVES  &14.21 & 42.09 & 32.67 & 26.54 & 19.11 & 86.11 & 43.52 & 78.33 \\
     & \method & \textbf{27.08} & \textbf{77.32} & \textbf{53.87} & \textbf{59.92} & \textbf{36.04} & \textbf{90.96} & \textbf{56.77} & \textbf{97.92} \\
\midrule
\multirow{7}{*}{mAP@0.8IoU} 
     & CRNN  &2.43 & 12.96 & 5.04 & 3.39 & 1.90 & 37.10 & 16.54 & 30.05 \\
     & Faster-RCNN &3.43&29.08&2.24&3.03&\textbf{9.53}& 53.54 &9.09& 69.06 \\
     & SEDT & 0.16 & 0.18 & 0.12 & 0.10 & 0.11 & 0.13 &0.16& 0.10 \\
     & Frame-ATST &4.72 & 18.20 & 10.55 & 8.28 & 2.66 & 24.70 & 23.44 & 27.11 \\
     & BEATs &5.18 & 18.98 & 10.31 & 9.77 & 2.93 & 51.00 & 27.11 & 42.27 \\
     & BirdAVES &4.52 & 19.28 & 9.12 & 5.14 & 3.08 & 48.29 & 23.04 & 42.44 \\
     & \method & \textbf{9.58} & \textbf{43.03} & \textbf{20.26} & \textbf{22.54} & 7.86 & \textbf{66.18} & \textbf{35.89} & \textbf{81.23} \\
\bottomrule
\end{tabular}
}
\caption{Mean average precision scores at 0.5 and 0.8 \ac{iou}. Best results in \textbf{bold}. With one exception, \method outperforms existing methods, and is sometimes far ahead, for example on BV10, HawB, and OZF. 
} 
\label{tab:main-results}
\end{table*}

As shown in Table \ref{tab:main-results}, \method outperforms other methods in almost all cases, and in is far ahead of all other models in several cases, e.g. 10+ points on mAP@0.5 on BV10,  HawB, HbW, and Katy. At mAP@0.8, \method scores 5+ points ahead of others on BV10, HawB, HbW, MT, and OZF. The diversity of animal sounds in the datasets especially highlights the general effectiveness of our method. Faster-RCNN generally performs well on  OZF and MT, and slightly surpasses \method on Katy mAP@0.8. However, it struggles with datasets with more than one class (AnSet, HawB, and Pow), as well as HbW. 
Of the frame-level \ac{sed} models, Frame-ATST, BEATs and BirdAVES, BEATs is generally the strongest, which is consistent with our findings for the backbone choice in \method (see Table \ref{tab:ablation-results}). SEDT is poor. Pretrained on datasets mostly of ambient city noises, it transfers badly to animal vocalizations. 



\subsection{Ablation Studies} \label{subsec:ablation-studies}

\begin{table*}
\tiny
\centering
\sisetup{
    reset-text-series = false, 
    text-series-to-math = true, 
    mode=text,
    tight-spacing=true,
    round-mode=places,
    round-precision=3,
    table-format=2.2,
    table-number-alignment=center
}
\resizebox{\textwidth}{!}{
\begin{tabular}{llcccccccc}
\toprule
Metric & Method & AnSet & BV10 & HawB & HbW & Katy & MT & Pow & OZF \\
\midrule
\multirow{3}{*}{mAP@0.5IoU} 
     & \method & \textbf{27.08} & \textbf{77.32} & \textbf{53.87} & \textbf{59.92} & \textbf{36.04} & \textbf{90.96} & \textbf{56.77} & \textbf{97.92} \\
     & with BirdAVES encoder & 22.86 & 46.33 & 49.22  & 48.04 & 26.59 & 88.78 & 50.21 & 96.36 \\
     & no fwd-bck matching & 25.04 & 75.97 & 52.10 &  56.99 & 34.97 & 89.39 & 50.02 & 95.77\\
\midrule
\multirow{3}{*}{mAP@0.8IoU} 
     & \method & \textbf{9.58} & \textbf{43.03} & \textbf{20.26} & \textbf{22.54} & \textbf{7.86} & \textbf{66.18} & \textbf{35.89} & \textbf{81.23} \\
     & w/ BirdAVES encoder& 8.64 & 25.56 & 18.55  & 12.22 & 5.16 & 56.31 & 32.31 & 74.32 \\
     & no fwd-bck matching & 7.46 & 37.70 & 16.82 & 18.73 & 7.14 & 36.22 & 26.79 & 80.36 \\
\bottomrule
\end{tabular}
}
\caption{Ablation studies on the backbone encoder and the forward-backward matching method. The main model uses the BEATs encoder. Best results in \textbf{bold}. Both ablation settings give a moderate, consistent drop in performance, showing the superiority of the BEATs encoder over BirdAVES, and the effectiveness of the \method forward-backward matching method. } \label{tab:ablation-results}
\end{table*}

Table \ref{tab:ablation-results} shows the effect of changing the encoder backbone of \method
, and of removing the forward-backward matching procedure. We found that using BirdAVES as a backbone for \method reduced performance compared with the version of that used the BEATs encoder. This was surprising considering BirdAVES was designed specifically for animal sounds; however differences in pre-training data volume and training regimes may explain the performance difference.
Removing forward-backward matching (i.e. only using forward predictions) also consistently lowers the mAP scores. Mostly the difference is 1-2 points but larger for some datasets, e.g. HbW and Pow.


\subsection{Performance on OZF-synthetic}

\begin{figure}
    \centering
    \includegraphics[width=\linewidth]{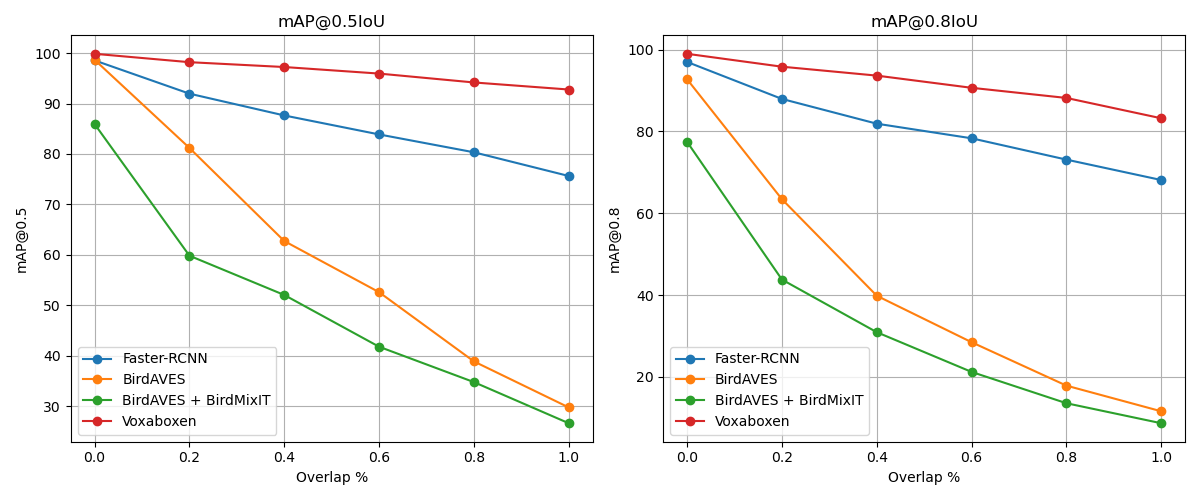}
    \caption{mAP on the synthetic portion of our OZF dataset, which comes in six varieties with increasing overlap, ranging from 0 to 1 in increments of 0.2. Existing methods, especially the frame-based BirdAVES and BirdAVES + BirdMixIT, deteroriate as the overlap ratio increases. \method is consistently the most accurate and drops only slightly with increasing overlap.}
    \label{fig:ozf-synth-results}
\end{figure}

Figure \ref{fig:ozf-synth-results} shows the performance of \method on the synthetic portion of our newly-released OZF dataset, for increasing overlap ratio. \method outperforms the comparison models at all overlap ratios, and maintains a high accuracy (92.77 mAP@0.5, 83.19 mAP@0.8), even at the highest over ratio of 1. BirdAVES, with and without BirdMixIT, deteriorates sharply with increasing overlap ratio, consistent with our argument that frame-based methods are insufficient for handling overlapping vocalizations. FasterRCNN fares better than other baselines, but is still firmly behind \method. 

\section{Conclusion}
In this work, we introduced \method, a novel method for sound event detection in bioacoustic recordings, specifically designed to address the challenges posed by overlapping vocalizations.  \method uses bidirectional predictions of vocalization boundaries combined with a graph-matching algorithm to accurately identify and localize events. To advance evaluation of overlapping vocalization detection, we released a new dataset, OZF, of zebra finch recordings with temporally-strong annotations and frequent overlaps. Extensive testing on seven existing datasets and our new dataset demonstrates that \method achieves state-of-the-art performance, with particularly notable improvements over standard \ac{sed} methods in scenarios with high overlap. This work highlights the potential of \method to advance bioacoustic research in ethology, ecology, and conservation.

\section*{Data Availability}
Data used in this study is available at \url{https://zenodo.org/records/15507508}. The code used in this study is available at \url{https://github.com/earthspecies/voxaboxen}.


\clearpage
\bibliographystyle{IEEEtran}
\bibliography{refs}

\newpage
\appendix

\section{OZF Further Details} \label{app:ozf-further-details}

\subsection{Segmentwise Statistics of the Real-World Portion}
\begin{figure*}[t!]
    \includegraphics[width=0.24\linewidth]{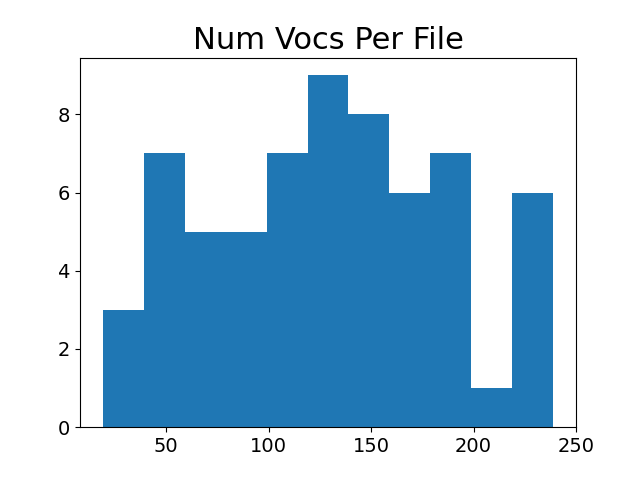} \hspace{-1em}
    \includegraphics[width=0.24\linewidth]{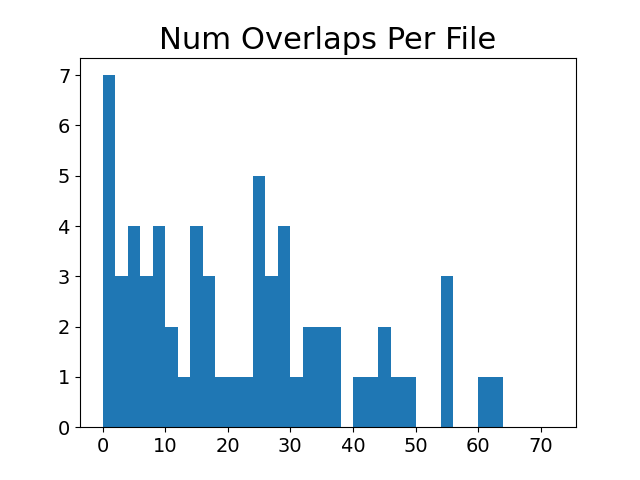}\hspace{-1em}
    \includegraphics[width=0.24\linewidth]{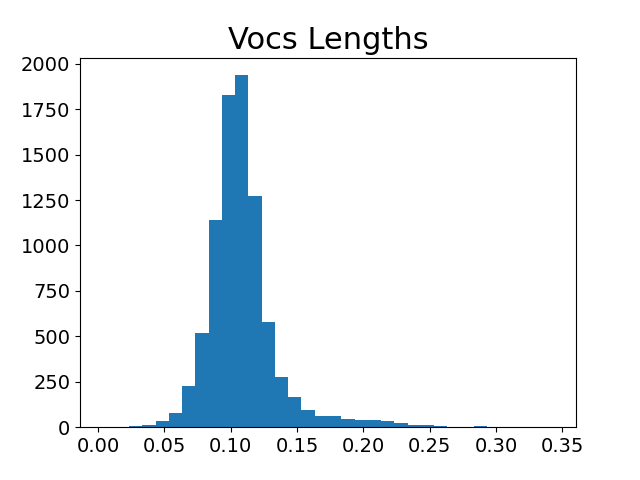}\hspace{-1em}
    \includegraphics[width=0.24\linewidth]{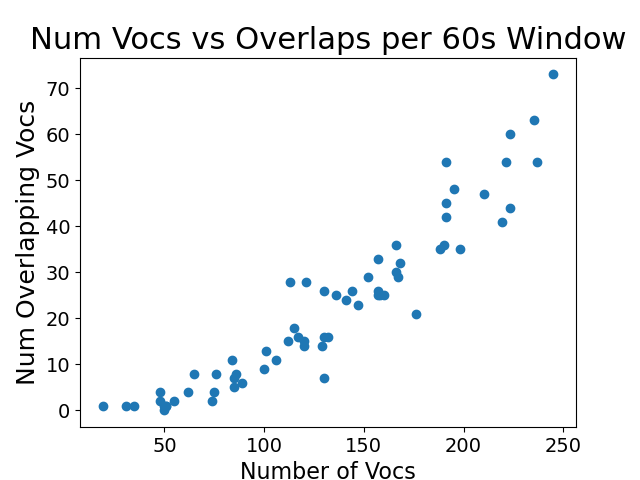}\hspace{-1em}
    \caption{Summary of the live portion of the dataset we release. Left: number of vocalizations per 60 s file. Second: number of overlapping pairs per 60 s file. Third: distribution of the lengths of vocalizations. Right: number of vocalizations per 60 s file vs number overlapping pairs.}
    \vspace{1em}
    \label{fig:dataset-stats}
\end{figure*}

\begin{figure*}[htb]
    \centering
    \includegraphics[width=0.3\linewidth]{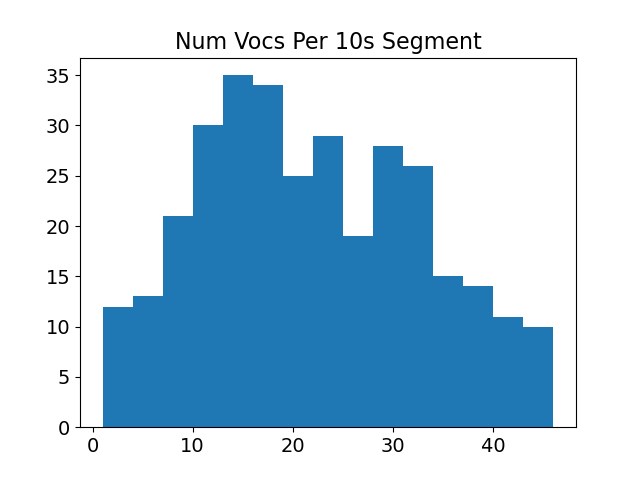}
    \includegraphics[width=0.3\linewidth]{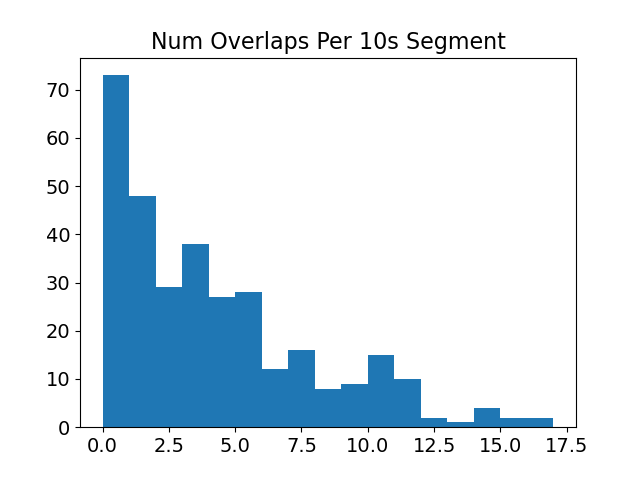}
    \includegraphics[width=0.3\linewidth]{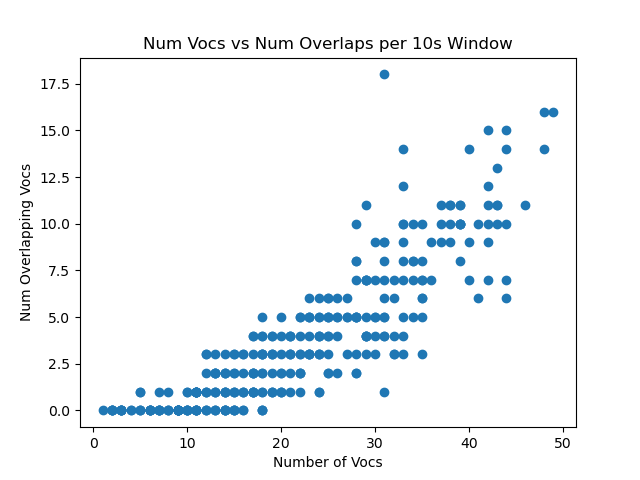}
    \caption{Statistics from the real-world portion of OZF, across 10 s segments. Top: the distribution of the number of vocalizations entirely contained within each 10 s segment, across all audio files. Middle: the distribution of the number of pairwise overlaps of these vocalizations. Bottom: the number of vocalizations vs the number of pairwise overlaps across each 10 s segment.}
    \label{fig:10s-dataset-stats}
\end{figure*}

Figure \ref{fig:dataset-stats} reports the distribution of vocalizations and overlaps across each 60-second audio file. Figure~\ref{fig:10s-dataset-stats} reports these per 10-second segment.

\subsection{Call Database for the Synthetic Portion} \label{app:call-database}

To construct the synthetic datasets, we created a database of female \ac{zf} calls. To construct this, female \acp{zf} were recorded using the same setup as with the live portion of OZF. Calls were detected using an initial version of \method, and cropped versions of the calls were saved. We then performed a denoising procedure: First, using BirdMixit~\cite{denton2022improving} each of these cropped calls was separated into four stems. Then, our trained model was again run across each of these four stems, and we retained a stem when \method both 1) detected a call and 2) the model detection confidence was higher on this stem than the other three stems. Finally, we observed that even after these steps there remained stems that contained no zebra finch vocalizations. To remove these, we performed a quality-filtering step: for each stem, we predicted the species of the call using BirdNET (Kahl et a., 2021). We retained only stems where BirdNET predicted a species with British English common name containing the word ``Finch''. The stems that passed this quality filter became the database of denoised female zebra finch calls.

\section{Expected Number of Overlaps from Independent Memoryless Sources} \label{app:expected-overlaps}

Given a window of time and some set $V$ of vocalizations whose onsets occur during this window, we are interested in the expected value of the number of pairs that overlap, assuming that the probability density function for the point of each onset is uniform and independent. Let $L$ be the length of the time window, 60s in our case, so that the pdf equals $\frac{1}{L}$. Then, for any two vocalizations with onsets as $v_1$ and $v_2$ of respective durations $l_1$ and $l_2$, the probability of overlap is 
\begin{equation} \label{eq:pairwise-overlap-prob}
    \frac{l_1 + l_2}{L}\,,
\end{equation}
because they will overlap if and only if $v_1$ falls in the interval $(v_2 - l_1, v_2+l_1)$, which is of length $l_1 + l_2$.

The expected number of overlaps, $\mathbb{E}[X]$, is the sum, across all ordered pairs of vocalizations, of the the indicator random variable for the event that they overlap, which equals the probability as given in \eqref{eq:pairwise-overlap-prob}. Let $|V|=n$, and let $l_i$ be the duration of the $i$ vocalization, then
\begin{align*}
    \mathbb{E}[X] =& \sum_{i=1}^n \sum_{j=1}^{i-1} \frac{l_i + l_j}{L} = \frac{1}{L}\left(\sum_{i=1}^n \sum_{j=1}^{i-1} l_i + l_j\right) \\
                  =& \frac{1}{L}\left(\sum_{i=1}^n (i-1)l_i + \sum_{i=1}^{n}\sum_{j=1}^{i} l_j \right) \\
                  =& \frac{1}{L}\left(\sum_{i=1}^n (i-1)l_i + \sum_{j=1}^{n-1}\sum_{i=j}^{n} l_j\right) \\
                  =& \frac{1}{L}\left(\sum_{i=1}^n (i-1)l_i + \sum_{j=1}^{n-1} (n-j) l_j\right) \\
                  =& \frac{1}{L}\left(\sum_{i=1}^n (i-1)l_i + \sum_{i=1}^{n-1} (n-i) l_i\right) \\
                  =& \frac{1}{L}\left(\sum_{i=1}^n (i-1)l_i + \sum_{i=1}^{n} (n-i) l_i\right) \\
                  =& \frac{1}{L}\left(\sum_{i=1}^n (i-1)l_i + (n-i) l_i\right) \\
                  =& \frac{1}{L}\left(\sum_{i=1}^n (n-1)l_i\right) \\
                  =& (n-1)\frac{\sum_{i=1}^n l_i}{L} \,.
\end{align*}
Thus, we see that the expected number of overlaps for uniform independent vocalizations is the product of two factors. The first is the vocalization density 
\begin{equation} \label{eq:d-def}
d = \frac{\sum_{i=1}^n l_i}{L}\,,
\end{equation}
which is the ratio between the length of all vocalizations played back to back and the duration of the time window in which they occur, or equivalently, the expected number of vocalizations occurring at any one point. The second factor is the number of vocalizations (minus 1). 

\begin{equation} \label{eq:uniform-expected-overlaps}
    \mathbb{E}[X] = d(n-1)\,,
\end{equation}
where $d$ is as in \eqref{eq:d-def}.

In the case of our released dataset, we must also account for the fact that there is a finite number of birds (eight), and overlaps can only occur between vocalizations from two different birds. For two given vocalizations, let $S$ be the event that they come from different birds, which has probability 
\begin{equation} \label{eq:ps}
P(S) = 1 - \frac{\sum_{j=1}^B (\sum_{i=1}^n \mathbbm{1}(b_i = j))^2}{n^2}\,,
\end{equation}
where $B$ is the number of birds and $b_i$ is the bird that produced the $i$th vocalization. Let $Z_j = \sum_{i=1}^n \mathbbm{1}(b_i = j)$ be the random variable counting the number of times the $j$th bird vocalises in a given time window (60s for our dataset). Assuming this distribution is the same for all birds, we can drop the subscript and just write $Z$. The expression in \eqref{eq:ps} is linear apart from the square on $Z$, so we have
\begin{align*}
P(S) & = 1 - \frac{\sum_{j=1}^B \mathbb{E}[Z^2]}{n^2}  = 1 - \frac{B \mathbb{E}[Z^2]}{n^2}\ \\
& =  1 - \frac{B (\mathbb{E}[Z]^2 + \text{Var}(Z))}{n^2}\,.
\end{align*}
If we model the vocalizations of each individual bird as a Poisson distribution, then we have 
\[
\mathbb{E}[Z] = \text{Var}(Z) = \lambda = \frac{n}{B}\,,
\]
giving
\[
P(S) = 1 - \frac{B ((\frac{n}{B})^2 + \frac{n}{B})}{n^2} = 1 - \frac{\frac{n^2}{B} + n}{n^2} = 1 - (\frac{1}{B} + \frac{1}{n})\,.
\]
The value from \eqref{eq:uniform-expected-overlaps} is then the probability of overlap between two vocalizations given they come from separate birds: $\mathbb{E}[X|S] = d(n-1)$, and the total probability of overlap is then 
\begin{equation}
    \mathbb{E}[X] = \mathbb{E}[X|S]P(S)= d(n-1) (1 - \frac{1}{B} - \frac{1}{n})\,.
    \label{eq:expected-n-overlaps}
\end{equation}

\subsection{Difference Between Expected and Observed Overlaps}
Table \ref{tab:expected-vs-observed-noverlaps} shows the observed number of pairwise overlaps per file, compared with the expected number from \eqref{eq:expected-n-overlaps}. The former is consistently lower than the latter. Indeed, looking at the `difference' column, we see it has mean 9.73, and standard deviation 9.05. We can model this difference as a normal distribution by the central limit theorem, as it is the sum of 8 independent samples from the distribution of a single bird. With 65 files, the estimated population standard deviation of this normal distribution is 
\[
\frac{9.05}{\sqrt{65-1}} = \frac{9.05}{8} = 1.13\,,
\]
so the $t$-value is $\frac{9.73}{1.13} = 8.61$. This is highly significant, as the significance threshold for 64 degrees of freedom is 3.23 at $99.9\%$ confidence. 

\begin{table*}[]
\centering
\small
\resizebox*{!}{0.95\textheight}{
\begin{tabular}{rrrrrrr}
\toprule
file & n & d & B & expected overlaps & observed overlaps & difference \\
\midrule
0 & 106 & 0.19 & 8 & 16.94 & 11 & 5.94 \\
1 & 117 & 0.22 & 8 & 21.80 & 16 & 5.80 \\
2 & 157 & 0.29 & 8 & 39.17 & 25 & 14.17 \\
3 & 191 & 0.36 & 8 & 59.25 & 42 & 17.25 \\
4 & 195 & 0.36 & 8 & 60.76 & 48 & 12.76 \\
5 & 221 & 0.38 & 8 & 73.36 & 54 & 19.36 \\
6 & 51 & 0.08 & 8 & 3.59 & 1 & 2.59 \\
7 & 160 & 0.30 & 8 & 41.91 & 25 & 16.91 \\
8 & 223 & 0.37 & 8 & 71.97 & 44 & 27.97 \\
9 & 19 & 0.03 & 8 & 0.48 & 1 & -0.52 \\
10 & 48 & 0.09 & 8 & 3.80 & 2 & 1.80 \\
11 & 31 & 0.06 & 8 & 1.40 & 1 & 0.40 \\
12 & 50 & 0.08 & 8 & 3.43 & 0 & 3.43 \\
13 & 147 & 0.28 & 8 & 34.87 & 23 & 11.87 \\
14 & 210 & 0.39 & 8 & 71.42 & 47 & 24.42 \\
15 & 191 & 0.36 & 8 & 59.05 & 45 & 14.05 \\
16 & 219 & 0.40 & 8 & 75.20 & 41 & 34.20 \\
17 & 237 & 0.41 & 8 & 85.24 & 54 & 31.24 \\
18 & 235 & 0.44 & 8 & 90.18 & 63 & 27.18 \\
19 & 51 & 0.08 & 8 & 3.60 & 1 & 2.60 \\
20 & 50 & 0.09 & 8 & 3.57 & 1 & 2.57 \\
21 & 85 & 0.15 & 8 & 11.14 & 7 & 4.14 \\
22 & 136 & 0.24 & 8 & 28.27 & 25 & 3.27 \\
23 & 141 & 0.23 & 8 & 28.35 & 24 & 4.35 \\
24 & 74 & 0.13 & 8 & 8.32 & 2 & 6.32 \\
25 & 166 & 0.33 & 8 & 47.87 & 30 & 17.87 \\
26 & 65 & 0.13 & 8 & 6.90 & 8 & -1.10 \\
27 & 223 & 0.39 & 8 & 74.50 & 60 & 14.50 \\
28 & 168 & 0.31 & 8 & 45.60 & 32 & 13.60 \\
29 & 158 & 0.29 & 8 & 39.44 & 25 & 14.44 \\
30 & 120 & 0.20 & 8 & 20.36 & 15 & 5.36 \\
31 & 84 & 0.16 & 8 & 11.16 & 11 & 0.16 \\
32 & 245 & 0.46 & 8 & 97.64 & 73 & 24.64 \\
33 & 191 & 0.37 & 8 & 60.43 & 54 & 6.43 \\
34 & 75 & 0.11 & 8 & 7.21 & 4 & 3.21 \\
35 & 130 & 0.26 & 8 & 28.77 & 7 & 21.77 \\
36 & 130 & 0.23 & 8 & 25.50 & 26 & -0.50 \\
37 & 86 & 0.14 & 8 & 10.15 & 8 & 2.15 \\
38 & 152 & 0.27 & 8 & 35.18 & 29 & 6.18 \\
39 & 55 & 0.10 & 8 & 4.40 & 2 & 2.40 \\
40 & 176 & 0.29 & 8 & 44.43 & 21 & 23.43 \\
41 & 115 & 0.18 & 8 & 18.08 & 18 & 0.08 \\
42 & 157 & 0.28 & 8 & 37.95 & 33 & 4.95 \\
43 & 35 & 0.04 & 8 & 1.29 & 1 & 0.29 \\
44 & 85 & 0.16 & 8 & 11.32 & 5 & 6.32 \\
45 & 89 & 0.16 & 8 & 12.18 & 6 & 6.18 \\
46 & 198 & 0.35 & 8 & 60.38 & 35 & 25.38 \\
47 & 113 & 0.25 & 8 & 24.49 & 28 & -3.51 \\
48 & 112 & 0.22 & 8 & 20.79 & 15 & 5.79 \\
49 & 101 & 0.18 & 8 & 15.48 & 13 & 2.48 \\
50 & 157 & 0.29 & 8 & 38.71 & 26 & 12.71 \\
51 & 144 & 0.26 & 8 & 31.71 & 26 & 5.71 \\
52 & 62 & 0.13 & 8 & 6.76 & 4 & 2.76 \\
53 & 167 & 0.32 & 8 & 46.38 & 29 & 17.38 \\
54 & 120 & 0.23 & 8 & 23.86 & 14 & 9.86 \\
55 & 76 & 0.15 & 8 & 9.99 & 8 & 1.99 \\
56 & 190 & 0.33 & 8 & 53.87 & 36 & 17.87 \\
57 & 130 & 0.20 & 8 & 22.82 & 16 & 6.82 \\
58 & 166 & 0.36 & 8 & 51.14 & 36 & 15.14 \\
59 & 121 & 0.25 & 8 & 26.36 & 28 & -1.64 \\
60 & 132 & 0.23 & 8 & 25.98 & 16 & 9.98 \\
61 & 48 & 0.08 & 8 & 3.26 & 4 & -0.74 \\
62 & 100 & 0.19 & 8 & 16.00 & 9 & 7.00 \\
63 & 129 & 0.23 & 8 & 25.27 & 14 & 11.27 \\
64 & 188 & 0.34 & 8 & 55.06 & 35 & 20.06 \\
\bottomrule
\end{tabular}
}

\caption{Comparison of the expected number of overlaps by equation \ref{eq:expected-n-overlaps} and the observed number of overlaps, by 60s file.}
    \label{tab:expected-vs-observed-noverlaps}
\end{table*}


\subsection{Evaluation Dataset Preprocessing} \label{app:dataset-preprocessing}

\textbf{AnuraSet} We used the portion of the frog call dataset presented in ~\cite{canas2023dataset} that includes strong temporal annotations (onset, offset, and species label). We randomly assigned files into train, validation, and test sets with ratios $60\%/20\%/20\%$. For our purpose, we retained only annotations corresponding to the ten most commonly occurring species in the dataset.

\textbf{BirdVox-10h} We used the version of the BirdVox dataset presented in~\cite{nolasco2023learning}. We divided each recording into three segments: the first $60\%$ was assigned to the train set, the next $20\%$ was assigned to the validation set, and the final $20\%$ was assigned to the test set. For our purpose, we merged all annotations (species labels for multiple passerine species) into a single class (passerine vocalization). This was done to avoid having many classes with few example vocalizations.

\textbf{Hawaiian Birds} We used the dataset of Hawaiian soundscapes presented in~\cite{amanda_navine_2022_7078499}. We randomly assigned files into train, validation, and test sets with ratios $60\%/20\%/20\%$. For our purpose, we retained only annotations corresponding to the nine most commonly occurring bird species in the dataset.

\textbf{Humpback} We used the ``initial'' audit portion of the dataset of humpback whale vocalizations presented in~\cite{humpbackdataset}, retaining only the 75-second clips containing at least one annotation. We randomly assigned these clips into train, validation, and test sets with ratios $60\%/20\%/20\%$. Finally, we retained only annotations corresponding to humpback whales, and discarded other annotations (e.g. ship noise).

\textbf{Katydid} We used the dataset of katydid calls presented in~\cite{madhusudhana2024extensive}. We randomly assign files into train, validation, and test sets with ratios $60\%/20\%/20\%$. For our purpose, we merged all annotations (species labels) into a single class (katydid call). This was done to avoid having many classes with few example calls.

\textbf{Meerkat} We used the dataset of on-body Meerkat recordings presented in~\cite{nolasco2023learning} (abbreviated as MT in \textit{loc. cit.}). We divided each recording into three segments: the first $60\%$ was assigned to the train set, the next $20\%$ was assigned to the validation set, and the final $20\%$ was assigned to the test set. For our purpose, we merged all annotations (vocalization type labels) into a single class (meerkat vocalization). This was done to avoid having many classes with few example vocalizations.

\textbf{Powdermill} We used the dataset of Northeastern United States soundscapes presented in~\cite{Chronister2021}. We randomly assigned files into train, validation, and test sets with ratios $60\%/20\%/20\%$. For our purpose, we retained only annotations corresponding to the six most commonly occurring bird species in the dataset.

\end{document}